\documentclass[fleqn,usenatbib]{mnras}

\usepackage{newtxtext,newtxmath}

\usepackage[T1]{fontenc}

\DeclareRobustCommand{\VAN}[3]{#2}
\let\VANthebibliography\thebibliography
\def\thebibliography{\DeclareRobustCommand{\VAN}[3]{##3}\VANthebibliography}

\usepackage{graphicx}	
\usepackage{amsmath}	
\usepackage{amsbsy}
\usepackage{ulem}
\usepackage[utf8]{inputenc}
\usepackage{scalerel}
\usepackage{tikz}
\usetikzlibrary{svg.path}

\definecolor{orcidlogocol}{HTML}{A6CE39}
\tikzset{
  orcidlogo/.pic={
    \fill[orcidlogocol] svg{M256,128c0,70.7-57.3,128-128,128C57.3,256,0,198.7,0,128C0,57.3,57.3,0,128,0C198.7,0,256,57.3,256,128z};
    \fill[white] svg{M86.3,186.2H70.9V79.1h15.4v48.4V186.2z}
                 svg{M108.9,79.1h41.6c39.6,0,57,28.3,57,53.6c0,27.5-21.5,53.6-56.8,53.6h-41.8V79.1z M124.3,172.4h24.5c34.9,0,42.9-26.5,42.9-39.7c0-21.5-13.7-39.7-43.7-39.7h-23.7V172.4z}
                 svg{M88.7,56.8c0,5.5-4.5,10.1-10.1,10.1c-5.6,0-10.1-4.6-10.1-10.1c0-5.6,4.5-10.1,10.1-10.1C84.2,46.7,88.7,51.3,88.7,56.8z};
  }
}

\newcommand\orcidicon[1]{\href{https://orcid.org/#1}{\mbox{\scalerel*{
\begin{tikzpicture}[xscale=1,yscale=-1, transform shape]
\pic{orcidlogo};
\end{tikzpicture}
}{|}}}}

\newcommand{\tess}{{\it TESS}}
\newcommand{\asassn}{{\it ASAS-SN}}

\newcommand{\gaia}{{\it Gaia}}
\newcommand{\xmm}{{\it XMM-Newton}}

\newcommand{\atlas}{{\it ATLAS}}
\newcommand{\opticam}{{OPTICAM}}
\newcommand{\blackgem}{{\it BlackGEM}}

\title[Micronova in DW Cnc]{DW Cnc: a micronova with a negative superhump and a flickering spin}

\author[M. Veresvarska et al.]{M. Veresvarska$^{\orcidicon{0000-0002-0146-3096}}$$^{1}$\thanks{E-mail: martina.veresvarska@durham.ac.uk},
S. Scaringi$^{\orcidicon{0000-0001-5387-7189}}$$^{1,2}$,
C. Littlefield$^{3}$,
D. de Martino$^{\orcidicon{0000-0002-5069-4202}}$$^{2}$,
C. Knigge$^{\orcidicon{0000-0002-1116-2553}}$$^{4}$,
J. Paice$^{\orcidicon{0000-0003-1149-1741}}$$^{1}$,
\newauthor
D. Altamirano$^{\orcidicon{0000-0002-3422-0074}}$$^{4}$,
A. Castro$^{\orcidicon{0000-0002-7832-5337}}$$^{5,4}$,
R. Michel$^{\orcidicon{0000-0003-1263-808X}}$$^{5}$,
N. Castro Segura$^{\orcidicon{0000-0002-5870-0443}}$$^{6}$,
J. Echevarr\'ia$^{\orcidicon{0000-0001-5960-3023}}$$^{7}$
P. J. Groot$^{\orcidicon{0000-0002-4488-726X}}$$^{8,9,10}$,
\newauthor
J. V. Hern\'andez Santisteban$^{\orcidicon{0000-0002-6733-5556}}$$^{11}$,
Z.A. Irving$^{\orcidicon{0009-0006-0951-3429}}$$^{4}$,
L. Altamirano-D\'evora$^{\orcidicon{0000-0001-7715-2182}}$$^{5,12}$,
A. Sahu$^{6}$,
\newauthor
D.A.H. Buckley$^{\orcidicon{0000-0002-7004-9956}}$$^{9,10,13}$,
F. Vincentelli$^{\orcidicon{0000-0002-1481-1870}}$$^{4}$.\\
$^{1}$Centre for Extragalactic Astronomy, Department of Physics, Durham University, South Road, Durham, DH1 3LE\\
$^{2}$INAF-Osservatorio Astronomico di Capodimonte, Salita Moiariello 16, I-80131 Naples, Italy\\
$^{3}$Bay Area Environmental Research Institute, Moffett Field, CA 94035, USA\\
$^{4}$School of Physics and Astronomy, University of Southampton, Highfield, Southampton SO17 1BJ, UK\\
$^{5}$Instituto de Astronom\'ia, Universidad Nacional Aut\'onoma de M\'exico, Carretera Tijuana-Ensenada Km. 107, Ensenada, B.C. 22860, M\'exico\\
$^{6}$Department of Physics, University of Warwick, Gibbet Hill Road, Coventry CV4 7AL, UK\\
$^{7}$Instituto de Astronom\'ia, Universidad Nacional Aut\'onoma
de M\'exico, Apartado Postal 70-264, Ciudad Universitaria, Ciudad de M\'exico, C. P. 04510, M\'exico\\
$^{9}$South African Astronomical Observatory, PO Box 9, Observatory 7935, Cape Town, South Africa\\
$^{8}$Department of Astronomy, University of Cape Town, Private Bag X3, Rondebosch 7701, South Africa\\
$^{10}$Department of Astrophysics/IMAPP, Radboud University, P.O. Box 9010, 6500 GL Nijmegen, The Netherlands\\
$^{11}$SUPA School of Physics \& Astronomy, University of St. Andrews, St. Andrews, UK\\
$^{12}$Facultad de Ingenieria, Arquitectura y Dise\~no, Universidad Aut\'onoma de Baja California, km. 103 Carretera Tijuana - Ensenada, C.P. 22860. M\'exico\\
$^{13}$Department of Physics, University of the Free State, P.O. Box 339, Bloemfontein 9300, South Africa\\
}

\date{Accepted XXX. Received YYY; in original form ZZZ}

\pubyear{\the\year{}}

\begin{document}
\label{firstpage}
\pagerange{\pageref{firstpage}--\pageref{lastpage}}
\maketitle

\begin{abstract}
Magnetic accreting white dwarfs in cataclysmic variables have been known to show bursts driven by different physical mechanisms; however, the burst occurrence is much rarer than in their non-magnetic counterparts. DW Cnc is a well-studied intermediate polar that showed a burst with a 4-magnitude amplitude in 2007. Here we report on a recent burst in DW Cnc observed by \asassn\ that reached a peak luminosity of 6.6 $\times$ 10$^{33}$ erg~s$^{-1}$, another 4 mag increase from its quiescent high state level. The released energy of the burst suggests that these are micronovae, a distinctive type of burst seen in magnetic systems that may be caused by a thermonuclear runaway in the confined accretion flow. Only a handful of systems, most of them intermediate polars, have a reported micronova bursts. We also report on the reappearance of the negative superhump of DW~Cnc as shown by \tess\ and \opticam\ data after the system emerges from its low state and immediately before the burst. We further report on a new phenomenon, where the spin signal turns "on" and "off" on the precession period associated with the negative superhump, which may indicate pole flipping. The new classification of DW Cnc as a micronova as well as the spin variability show the importance of both monitoring known micronova systems and systematic searches for more similar bursts, to limit reliance on serendipitous discoveries. 
\end{abstract}




\section{Introduction}
\label{s:intro}

Cataclysmic variables (CVs) are the most common type of accreting white dwarfs (AWDs). They consist of a binary system in which a white dwarf (WD) accretes mass from a low-mass star via Roche lobe overflow. For an in-depth review of CVs and their history see \citet{Warner1995} and \citet{Knigge2011}.

In general, CVs can be divided into two subcategories based on their magnetic field strength. CVs whose magnetic field is too weak to disrupt the accretion flow ($\lesssim 10^{6}$ G) are referred to as "non-magnetic". On the other hand systems with magnetic field strength above $\sim 10^{6}$ G are referred to as magnetic. Depending on the WD magnetic field strength only an outer accretion disc may form. The inner parts will be inhibited by the Alfven radius, i.e. they will be truncated at the magnetospheric radius, where the matter attaches to magnetic field lines and is accreted onto the magnetic poles of the WD via accretion columns. In these systems, the spin period of the WD is not synchronized with the orbital period; they generally display a periodic modulation on the WD spin frequency associated with emission from the accretion column. Such systems are known as Intermediate Polars (IPs), where the matter is accreted via an arc-shaped curtain \citep{Rosen1988MNRAS.231..549R}. For larger magnetic fields the magnetospheric radius can lie outside of the disc co-rotation radius, which inhibits entirely the formation of an accretion disc. In majority of these systems, the WD spin and binary orbit are synchronised and are commonly known as Polars, with the exception of rare discless IPs such as V2400 Oph \citep{Buckley1995MNRAS.275.1028B}.

AWDs display various types of transient-like brightness increases. The most common and well-established bursts in CVs are dwarf nova outbursts, in which a sudden increase in accretion rate, thought to be driven by thermal-viscous disc instabilities \citep{Osaki1996PASP..108...39O,Lasota2001NewAR..45..449L,Dubus2018}, causes the temperature and hence luminosity of the disc to increase. These outbursts usually last several days to months, with longer orbital period systems tending to have longer outbursts due to their larger discs. Once a cooling wave due to a lower ionisation level passes through and stops the outburst, the temperature of the disc will return to its quiescent level \citep{Lasota2001NewAR..45..449L}. The outbursts occur semi-periodically with some special cases where they increase in luminosity amplitude, leading up to a superoutburst. Whereas dwarf nova outbursts are very common in non-magnetic AWDs, they are extremely rare in magnetic systems \citep{Hameury2017A&A...602A.102H}.

A different type of burst exhibited by AWDs is the so-called magnetic gated bursts, which appear only in systems with low magnetic field WDs \citep{Scaringi2017}. In these systems, the material is accumulated at the edge of a truncated inner disc close to the WD surface, where the spinning material prevents it from accreting freely. When the disc pressure exceeds the pressure of the spinning magnetosphere matter can then accrete onto the WD. Several systems have now shown magnetically gated bursts (MV Lyrae: \citealt{Scaringi2017}, TW Pic: \citealt{Scaringi2021}, V1233 Sgr: \citealt{Hameury2022}, V1025 Cen: \citealt{littlefield22}). 

A very different kind of burst is displayed by a growing number of systems. They are fast ($\gtrsim$ 1 day), bright, and somewhat isolated bursts which cannot be fully reconciled with magnetic gating. One example is TV Col, which has been shown to display fast outflows during burst maximum \citep{SM84}. These bursts, which appear phenomenologically different from all of the aforementioned types of bursts, are usually referred to as micrononvae \citep{Scaringi2022}. One currently unproven hypothesis for the origin of these bursts is that they are the result of a localised thermonuclear runaway occurring on the WD magnetic poles\citep{Scaringi2022a}; however their true physical origin is yet to be unambiguously determined. These bursts can last from $\sim$ 10 hours up to a few days. To date, there have been very few targets reported to show these bursts \citep{Scaringi2022,Veresvarska2024,Ilkiewicz2024ApJ...962L..34I,Irving2024MNRAS.530.3974I}. However, with the advent of new synoptic sky surveys \citep[e.g., \blackgem{};][]{blackgem}, the number of detections is expected to rise allowing for a better population characterisation of these systems.

\citet{Ilkiewicz2024ApJ...962L..34I} introduced a convenient way to distinguish between the observational characteristics of the various types of bursts observed in AWDs via diagnostic diagrams. The diagrams plot the burst energies, peak luminosities, durations, and recurrence times of various kinds of CV outbursts.  They show that dwarf nova outbursts, micronovae, and magnetically-gated bursts occupy distinct regions of parameter
space, suggesting that these phenomena all arise from distinct physical mechanisms.


DW Cnc is a well studied IP \citep{Stepanyan1982PZ.....21..691S,Rodriguez-Gil2004MNRAS.349..367R} which is known to undergo periods of low states as also observed in other IPs \citep{Duffy2022MNRAS.510.1002D,Covington2022ApJ...928..164C}. It has a reported spectroscopic orbital period of 86.1 minutes \citep{Patterson2004PASP..116..516P,Rodriguez-Gil2004MNRAS.349..367R} and a 38.6 minute spin \citep{Rodriguez-Gil2004MNRAS.349..367R}. The spin signal has been known to disappear during the low state \citep{Montero2020MNRAS.494.4110S,Covington2022ApJ...928..164C}, only to emerge after the system recovers to its pre-low state levels \citep{Ramirez+2022,Duffy2022MNRAS.510.1002D}. There is also evidence of both hydrogen and helium in the accretion disc of DW Cnc; \citet{Montero2020MNRAS.494.4110S} reports double peaked H$\alpha$ and He $\lambda$5876 $\AA$ lines with the latter showing a spiral structure in the disc. The authors associate this feature to a similar behaviour seen on several other systems including IP Peg, U Gem, EC 21178-5417 and the IP system DQ Her \citep{Steeghs1997MNRAS.290L..28S,Groot2001ApJ...551L..89G,Bloemen2010MNRAS.407.1903B,Ruiz-Carmona2020MNRAS.491..344R} where such structures are considered to signify an enlarged accretion disc whose outer regions are being affected by the tidal forces of the donor e.g. \citet{Steeghs1997MNRAS.290L..28S}. Furthermore, DW Cnc has been known to show short duration ($<$1 day) bursts \citep{Crawford2008JAVSO..36...60C}. These bursts detected in 2007 correspond to a brightening of $\sim$ 4 mag. Disc instabilities and mass-transfer enhancements have been suggested as causes \citep{Crawford2008JAVSO..36...60C}, but their nature remains unknown.

Here we present new \textit{Transiting Exoplanet Satellite Survey} (\tess\,) photometric observations of DW Cnc showing 2 coherent periods that persist throughout the \tess\ monitoring. We associate these coherent periods with the spin and beat between the WD spin and the orbital period of the system. We present these results in Section \ref{ss:tess} along with the appearance of a negative superhump in \tess\ sectors 71 and 72 accompanied by variable spin in Section \ref{ss:spin}. We further report a burst observed in \textit{All-Sky Automated Survey for Supernovae } (\asassn\,) coverage of DW Cnc reaching a maximum observed luminosity of $\sim6.6 \times 10^{33}$ erg~s$^{-1}$. We present these results in Section \ref{ss:bursts} and their interpretation as micronova in Section \ref{ss:micronova}.

\section{Observations}
\label{s:obs}

In this section, we discuss the data used for this work. \tess\ data and its calibration with \asassn\ and \atlas\ is discussed in Section \ref{ss:tess_obs}, while the ground-based \opticam\ data is detailed in Section \ref{ss:opticam}.

\begin{figure*}
	\includegraphics[width=\textwidth]{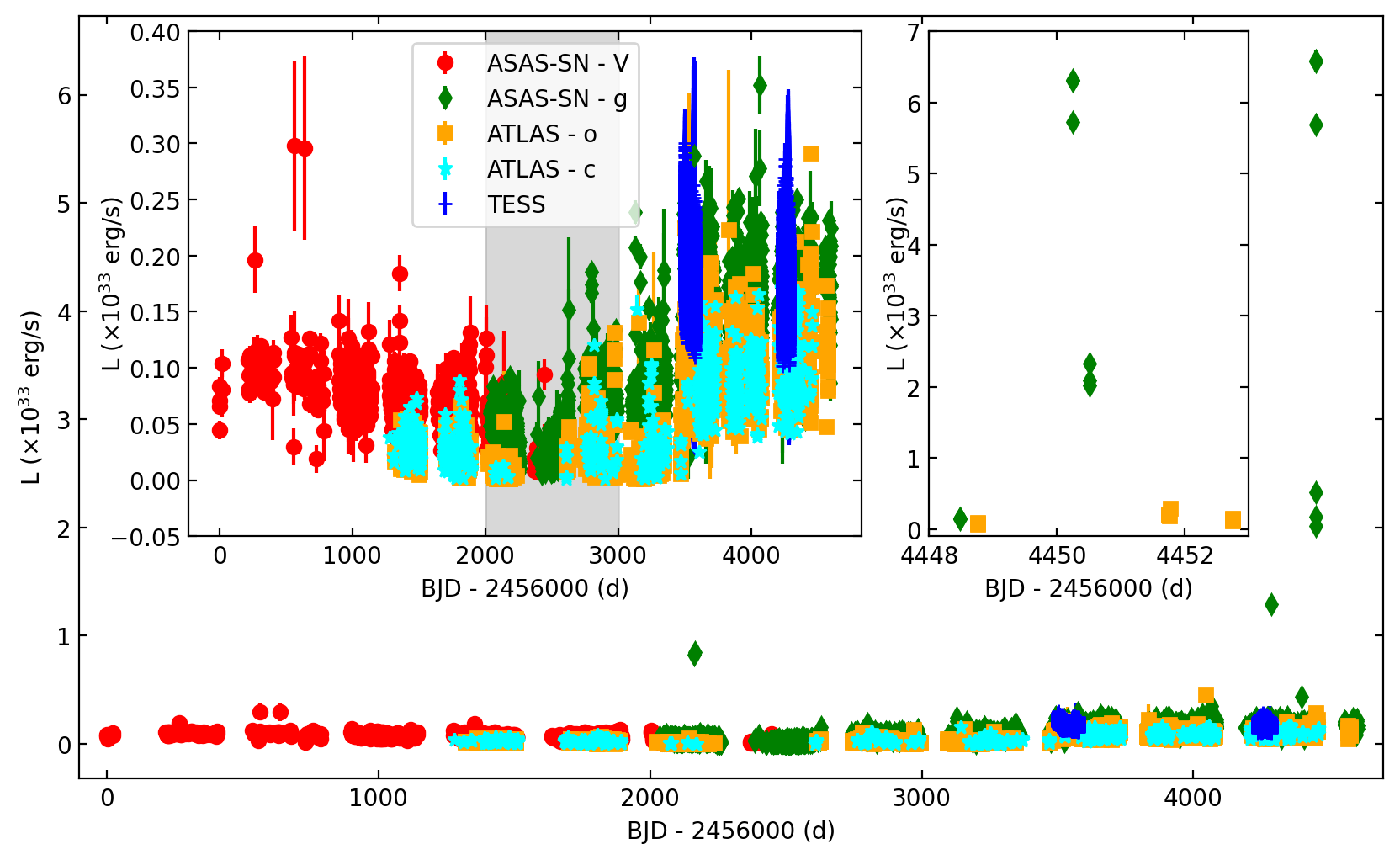}
    \caption{\asassn{} \atlas{} and \tess{} light curve of DW Cnc. Circles and diamonds denote the \asassn\ V and g bands respectively. Similarly, the \atlas{} o and c bands are denoted by squares and stars respectively. The crosses correspond to \tess\ sectors 44, 45, 46, 71 and 72. The left inset panel shows an enlargement in the luminosity axes, with the low state being marked by the shaded region. The right inset panel shows a zoom-in on the burst with $\sim6.6 \times 10^{33}$ erg~s$^{-1}$ peak observed luminosity.}
    \label{fig:LC}
\end{figure*}

\subsection{\tess\ and simultaneous ground-based \asassn\ and \atlas\ observations}
\label{ss:tess_obs}

The data analysed in this work were obtained from \tess\ and \asassn. \tess\ observed DW Cnc over five sectors from October 12$^{\rm{th}}$, 2021, to December 30$^{\rm{th}}$, 2021 (sectors 44, 45 and 46), and from October 16$^{\rm{th}}$, 2023, to December 7$^{\rm{th}}$, 2023 (sectors 71 and 72). \asassn\ observed DW Cnc from February 16$^{\rm{th}}$, 2012, to November 29$^{\rm{th}}$, 2018, in the V band, and from October 7$^{\rm{th}}$, 2017, to May 28$^{\rm{th}}$, 2024, in the $g$ band. All relevant observations are displayed in Figure~\ref{fig:LC}.

\tess\ data were downloaded, and cosmic rays were removed using the \texttt{Lightkurve} package\footnote{\url{https://github.com/lightkurve/lightkurve}} \citep{lightkurve2018ascl.soft12013L}. The Simple Aperture Photometry (SAP) flux was used to preserve the intrinsic variability of the systems, avoiding the transit detection optimization of the Pre-search Data Conditioning (PDCSAP) flux. Data points were excluded if their quality flag was greater than 0.

While \tess\ provides excellent relative photometric precision, data from another observatory are needed to achieve absolute photometry from \tess\ data. To convert \tess\ data to erg s$^{-1}$, it is necessary to consider ground-based observations and the distance to the target. Quasi-simultaneous \asassn\ $g$-band data \citep{Shappee2014, Kochanek_2017} were used for this purpose similarly to \citet{Scaringi2012a} and \citet{Veresvarska2024}. The $g$-band is centred at 475 nm with a width of 140 nm, partially overlapping with the \tess\ band-pass (600 -- 1000 nm). Despite the non-exact overlap of the passbands, we assume no colour-term variations. The data were obtained from the \asassn\ webpage\footnote{\label{asassn}\url{https://asas-sn.osu.edu/}}. The conversion was performed using simultaneous observations for each half sector. Simultaneous data include all \asassn\ observations within a \tess\ cadence of 2 minutes. Assuming a linear relation between the two bands, a direct conversion can be established as $F_{\rm ASAS-SN} \left[ \rm mJy \right] = A \times F_{ \rm TESS} \left[ \rm e^{-}s^{-1} \right]  + C$. The coefficients for each half-sector, accounting for data deviations due to the gap in the middle of \tess\ sectors, are specified in Table \ref{tab:cal}.

Figure \ref{fig:LC} shows the luminosity corresponding to the \tess\ and \asassn\ data corrected for the distance of $209 \pm 2$ pc inferred from \gaia\ DR3 parallax \citep{Gaia2023}. Additionally, Figure \ref{fig:LC} shows the \atlas\ o and c band forced-photometry light curve \citep{Tonry2018PASP..130f4505T,Heinze2018AJ....156..241H,Smith2020PASP..132h5002S}, corrected for the distance inferred from \gaia\ DR3 parallax. The \atlas\ forced photometry was obtained through the web interface \citep{Shingles2021TNSAN...7....1S}\footnote{\label{atlas}\url{https://fallingstar-data.com/forcedphot/}}. Combined \atlas\ and \asassn\ photometry improve the sampling of the observations and constraint on transient events not detected by only one survey (see right inset panel in Figure \ref{fig:LC}).

\subsection{\opticam\,}
\label{ss:opticam}

High-time resolution, triple-band optical observations with \opticam\,\footnote{\url{https://www.southampton.ac.uk/opticam}} \citep{Castro+19,Castro+24} were conducted using the 2.1 m telescope at the Observatorio Astron\'omico Nacional San Pedro M\'artir (OAN-SPM). 
The log of observations made with \opticam{} is shown in Table \ref{tab:opticam_log}. All observations carried out with \opticam{} used the filters $g'$, $r'$, and $i'$ simultaneously, with an exposure time of 20 seconds. Data were reduced following standard procedures using various tasks in NOIRLab IRAF v2.18\footnote{https://iraf.noirlab.edu/}~\citep{Fitzpatrick+24}. Star J075900.5+161647.1\footnote{also SDSS J075900.5+161646.5 with g=15.47; r=15.01 and i=14.84} was used as a reference. According to the GAIA synthetic photometry catalogue, this star has magnitudes $g'=15.32, r'=14.78$ and $i'=14.62$ \citep{Gaia2020yCat.1350....0G}.

\begin{table}
    \centering
    \caption{Log of multi-band \opticam{} observations. All observations presented in the table were carried out using the filters g', r', and i' simultaneously, with an exposure time of 20 seconds.}
    \label{tab:opticam_log}
    \begin{tabular}{lcc}
    \hline
    Date (UT) & HJD (start time) & Effective Exposure (h)\\
    \hline
     2022-12-10  & 2459924.8052 & 2.22 \\
     2022-12-15  & 2459928.8021 & 2.84 \\
     2023-02-05  & 2459980.6460 & 4.58 \\
     2023-02-06  & 2459981.8040 & 4.59 \\
     2023-02-10  & 2459985.7215 & 5.58 \\
     2023-02-14  & 2459989.6513 & 7.42 \\
     2024-01-10  & 2460319.6720 & 8.02 \\
     2024-01-12  & 2460321.7719 & 6.00 \\
     2024-01-14  & 2460323.6973 & 7.56 \\
     \hline
    \end{tabular}

\end{table}

\begin{table}
	\centering
	\caption{Summary of the conversion coefficients from \tess\ flux in $\rm e^{-}s^{-1}$} to \asassn\ flux in $\rm mJy$ for DW Cnc. As the conversion is done twice per \tess\ sector, all corresponding coefficients are listed.
	\label{tab:cal}
	\begin{tabular}{lccr} 
		\hline
		Sector & Sector half & $A$ $\bigg( \frac{\rm mJy}{\rm e^{-}s^{-1}} \bigg)$ & C $(\rm mJy)$ \\
		\hline
		44 & 1 & 0.021 $\pm$ 0.002 & $-$0.5 $\pm$ 0.4\\
          & 2 & 0.021 $\pm$ 0.002 & $-$0.5 $\pm$ 0.4\\
		45 & 1 & 0.035 $\pm$ 0.004 & 0.9 $\pm$ 0.4\\
		   & 2 & 0.036 $\pm$ 0.004 & 0.8 $\pm$ 0.4\\
        46 & 1 & 0.027 $\pm$ 0.003 & $-$0.4 $\pm$ 0.5\\
           & 2 & 0.03 $\pm$ 0.01 & 0 $\pm$ 2\\
        71 & 1 & 0.032 $\pm$ 0.001 & $-$0.2 $\pm$ 0.2\\
           & 2 & 0.032 $\pm$ 0.001 & $-$0.2 $\pm$ 0.2\\
        72 & 1 & 0.031 $\pm$ 0.002 & $-$0.4 $\pm$ 0.3\\
           & 2 & 0.032 $\pm$ 0.009 & $-$1 $\pm$ 2\\
		\hline
	\end{tabular}
\end{table}

\section{Results}
\label{s:res}

In Section \ref{ss:tess} we report all the coherent signals found in new \tess\ photometry of DW Cnc. In Section \ref{ss:spin} the variable nature of the spin signal of DW Cnc as seen in \tess\ is shown. We further report on the burst of DW Cnc observed by \asassn\ on the 19$^{\rm{th}}$ of May 2024 in Section \ref{ss:bursts}.

\subsection{\tess\ Data Analysis}
\label{ss:tess}

\begin{figure*}
	\includegraphics[width=\textwidth]{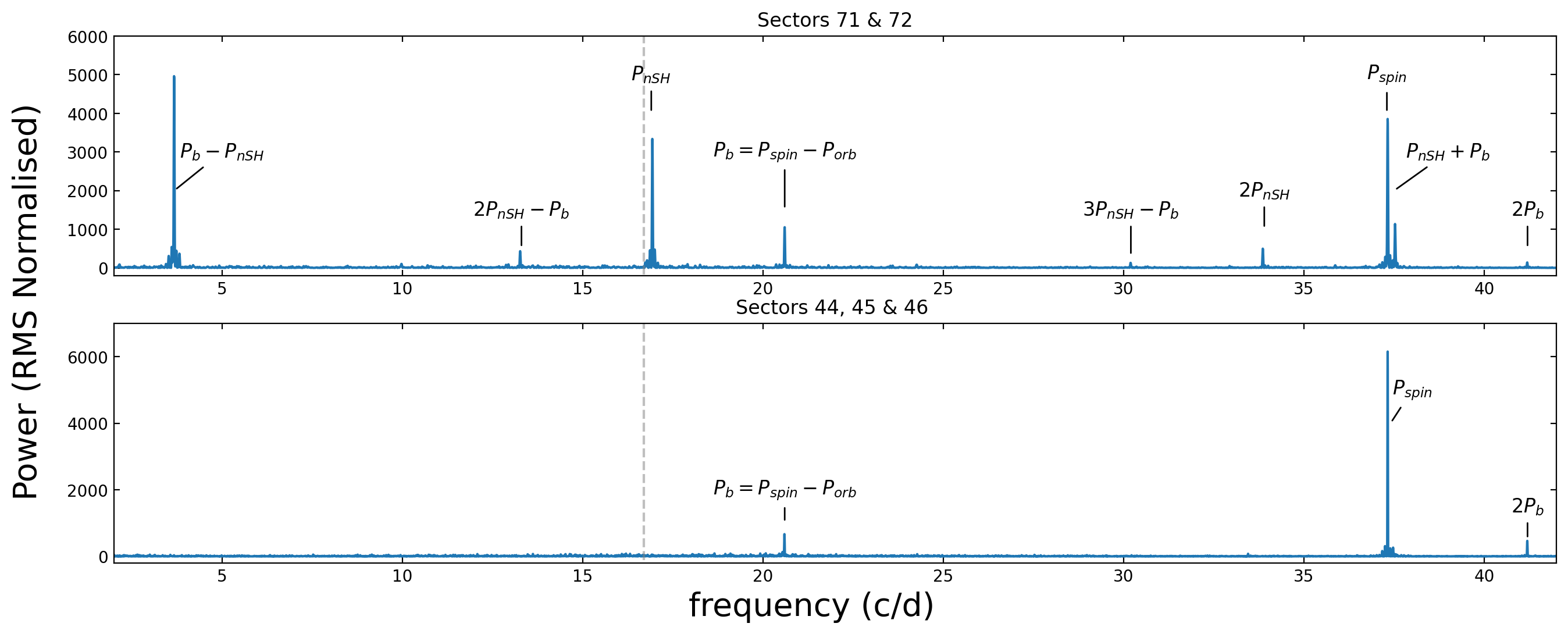}
    \caption{Lomb-Scargle power spectra of \tess\ sectors of DW Cnc zoomed in on all identified signals. The top panel shows the periodogram of the combined light curves of sectors 71 and 72 whilst the bottom panel shows the periodogram for the combined light curve of sectors 44 to 46. The position of the orbital period, which is not detected here, is marked by a dashed line.}
    \label{fig:PSD}
\end{figure*}

The \tess\ light curve of DW Cnc as described in Section \ref{s:obs} can be divided into 2 semi-continuous sections. One represents sectors 44 to 46 immediately after DW Cnc returned from a low state and before any of the reported bursts in Section \ref{ss:bursts} took place. The Lomb-Scargle \citep{Lomb1976} periodogram of this $\sim$ 3-month light curve is shown in the bottom panel of Figure \ref{fig:PSD}. In all instances in this work, the Lomb-Scargle implementation used is \texttt{Astropy} v.5.3.4, with the high limiting frequency corresponding to the Nyquist frequency of the given time series and the low frequency to the $\frac{3}{L}$, where $L$ represents the length of the time series. There are only 3 signals recovered from this light curve at 37.3 c d$^{-1}$, 20.6 c d$^{-1}$ and 41.2 c d$^{-1}$ and they are listed in Table \ref{tab:PSD}.

The other semi-continuous light curve is constructed from sectors 71 and 72, immediately before the small burst at $\sim$ 2460287 BJD. The Lomb-Scargle periodogram of the light curve is shown in the top panel of Figure \ref{fig:PSD}. The periodogram also shows all the signals present in sectors 44 through 46. However, on top of these signals, numerous new signals appear. All the signals are listed in Table \ref{tab:PSD} with their corresponding errors and sectors in which they are present.

The errors in Table \ref{tab:PSD} are determined via bootstrapping of the original light curve, similar to \citet{Paice2024MNRAS.531L..82P}. This consists of randomly selecting with replacement \(N\) number of data points from the light curve, where \(N\) is the original number of data points. Then the same Lomb-Scargle periodogram is constructed and the peak of the signal is extracted. A distribution of the peaks after 5 $\times$ 10$^{4}$ repetitions is constructed. A Gaussian fit to the resulting distribution gives the mean value of the signal and its error corresponding to a 1$\sigma$ deviation of the distribution.

\begin{table}
	\centering
	\caption{List of all coherent signals extracted from \tess\ data of DW Cnc and their relation to one another with the sectors in which they are present. The spin period is noted as $P_{\rm spin}$. The beat between the spin period and the undetected orbital period ($P_{\rm orb} = 16.7245$ c d$^{-1}$) is noted as $P_{\rm b}$ with its first harmonic $2P_{\rm b}$. The negative superhump associated with the retrograde nodal precession of a tilted accretion disc is noted as $P_{\rm nSH}$ with its first harmonic $2P_{\rm nSH}$. The beats between these frequencies and their harmonics are denoted by their combination. }
	\label{tab:PSD}
	\begin{tabular}{ccc} 
		\hline
		Frequency (c d$^{-1}$) & Origin & \tess\ sectors \\
		\hline
        3.66860(5) & $P_{\rm b} - P_{\rm nSH}$ & 71 \& 72 \\
        13.2602(7) & $2P_{\rm nSH} - P_{\rm b}$ & 71 \& 72 \\
        16.9276(2) & $P_{\rm nSH}$ & 71 \& 72 \\
        20.5962(2) & $P_{\rm b} = P_{\rm spin} - P_{\rm orb}$ & all \\
        30.189(1) & $3P_{\rm nSH} - P_{\rm b}$ & 71 \& 72 \\
        33.85669(8) & $2P_{\rm nSH}$ & 71 \& 72 \\
        37.32166(7) & $P_{\rm spin}$ & all \\
        37.52541(3) & $P_{\rm nSH} + P_{\rm b}$ & 71 \& 72 \\
        41.2024(2) & $2P_{\rm b}$ & all \\
        \hline
	\end{tabular}
\end{table}

\subsection{Spin Variability}
\label{ss:spin}

\begin{figure*}
	\includegraphics[width=1.0\textwidth,trim={0 3cm 0 0}]{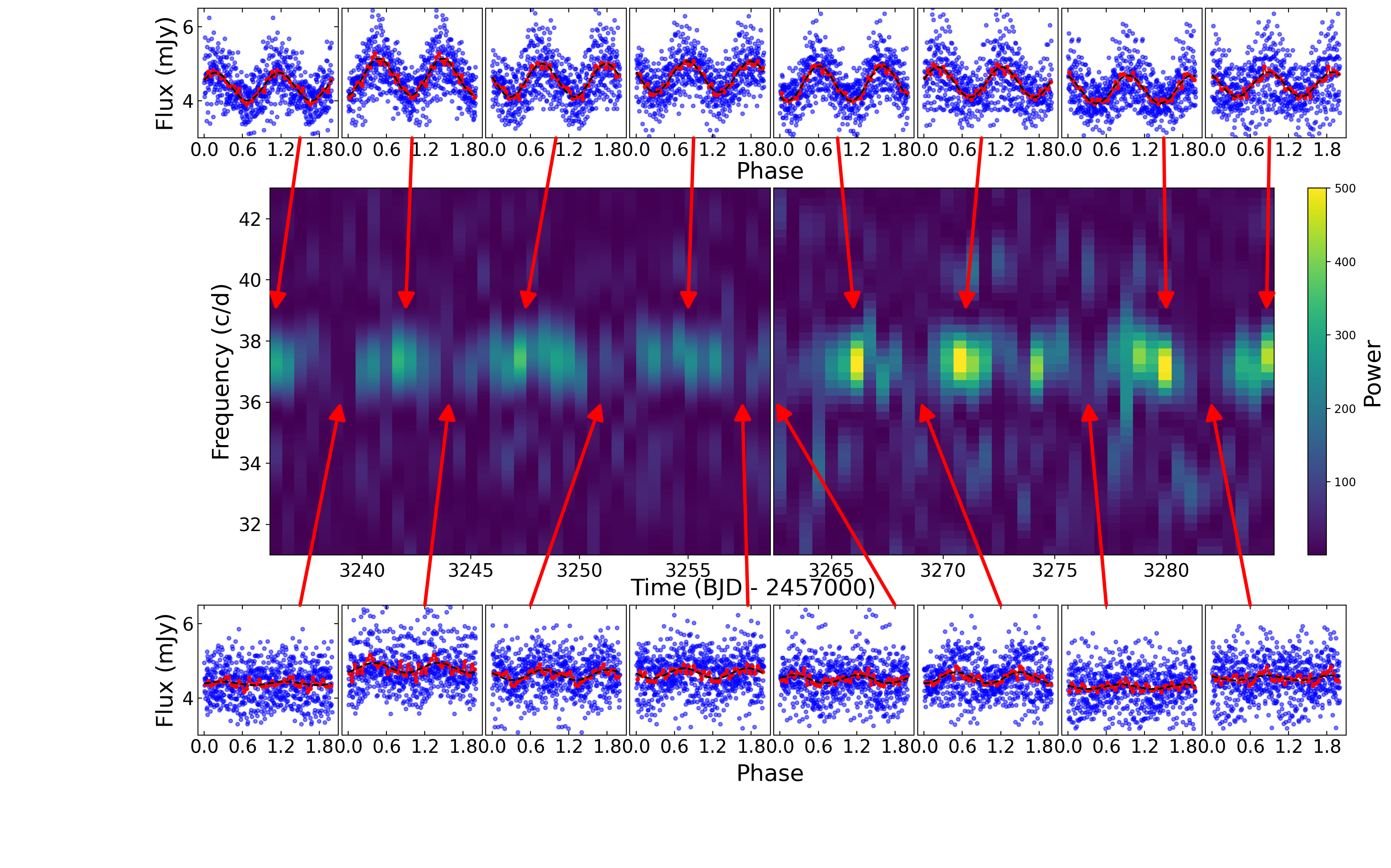}
    \caption{Dynamical Lomb-Scargle power spectra of \tess\ sectors 71 and 72 of DW Cnc zoomed in on the spin signal at 37.32166(7) c d$^{-1}$ showing semi-periodic turning ``on'' and ``off'' of itself. The top and bottom panels show a phase folded light curve on the spin period of 0.5 \(d\) long segment as indicated by the arrows. The top panels are selected at the peak spin power in a given cycle on the spin being "on". The bottom panels show the minimum power segment in the "off" segment.}
    \label{fig:dynPSD}
\end{figure*}

Figure \ref{fig:dynPSD} shows the dynamical power spectral density (PSD) of \tess\ sectors 71 and 72, with zoom-ins showing the light curves of several sections folded on the spin period from Table \ref{tab:PSD}. The dynamical PSD is constructed by taking Lomb-Scargle periodograms, as in Section \ref{ss:tess}, of 0.5-day long non-overlapping sections of the light curve. While a strong signal is present throughout the 2 sectors, there is also a significant time-dependent variability, with 2 distinct states, that are not seen in the previous \tess\ sectors 44$-$46: An "on" state, where the spin signal is showing significant power and lasts $\sim$ 3 days centred around BJD $-$ 2457000 3235, 3242, 3247, 3255, 3267, 3272, 3279 and 3285; and an "off" state in between these, when the power at $P_{\rm spin}$ drops significantly to the background level of Poisson noise $P_{\rm RMS~normalised}\lesssim$ 200. This latter state is much shorter ($\sim$ 1 day) in duration and occurs around BJD - 2457000 3238, 3244, 3252, 3258, 3260, 3270, 3277 and 3282. To demonstrate the change in spin state, examples of phase-folded light curves are provided for the ``on'' (top panels of Figure \ref{fig:dynPSD}) and ``off" (bottom panels of Figure \ref{fig:dynPSD}) states. The examples of the phase-folded light curves correspond to the 0.5-day segments of the light curve at instances indicated by the red arrows. As seen from the phase-folded light curves, the flux level remains constant in both states and there is no significant change in the spin period $P_{\rm spin}$ detected over the course of the \tess\ sectors with respect to the ephemeris reported by \citet{Patterson2004PASP..116..516P}. As seen in the spin pulse profiles in Figure \ref{fig:dynPSD} only shows 1 peak, which is also true of other signals detected in Sectors 71 and 72. However, when examining the shape of the spin-orbit beat pulse profile over the course of the negative superhump period from Table \ref{tab:PSD} shows a $\sim180^{\circ}$ phase shift. 

\begin{figure*}
	\includegraphics[width=1\textwidth,trim={5cm 0 5cm 0}]{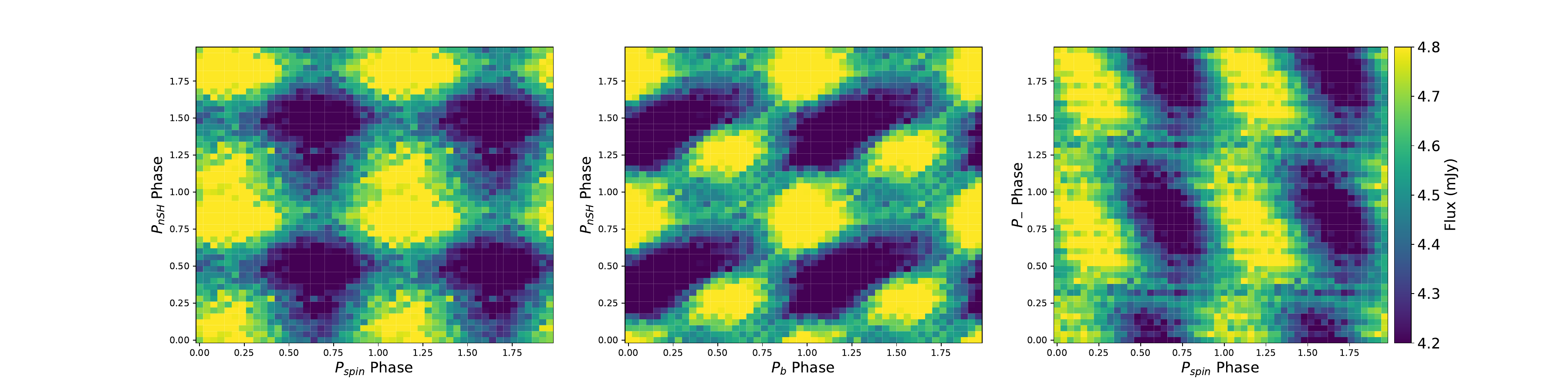}
    \caption{2D profiles of the spin and beat pulses across the superhump cycle and spin profile across the fundamental precession cycle. The beat profile shows a $\sim180^{\circ}$ jump for half of the superhump cycle similarly to the $\sim180^{\circ}$ phase shift in spin phase over the precession cycle, indicating that the tilt of the disk determines which magnetic pole is dominant.}
    \label{fig:2dprofiles}
\end{figure*}

This is demonstrated in Figure \ref{fig:2dprofiles}, where the spin period (left hand-side) and spin-orbit beat period (centre) pulse profiles are shown as a function of the negative superhump cycle, similarly to \citet{Littlefield2019ApJ...881..141L,Littlefield2021AJ....162...49L,Littlefield2023AJ....165...43L}. Figure \ref{fig:2dprofiles} demonstrates a dynamical representation of phase folded light curves on a signal at different phases of another signal. This allows us to identify the changes in positions of the peaks and troughs of the phase-folded light curves on a given period with respect to the phase of another one. Hence, a diagonal pattern signifies a phase shift between the given signals, as seen in the middle and right panel of Figure \ref{fig:2dprofiles}. Furthermore, the $\sim180^{\circ}$ phase drift in spin cycle is also visible during the precession cycle associated with the negative superhump identified in Table \ref{tab:PSD}.

\begin{figure}
	\includegraphics[width=1\columnwidth]{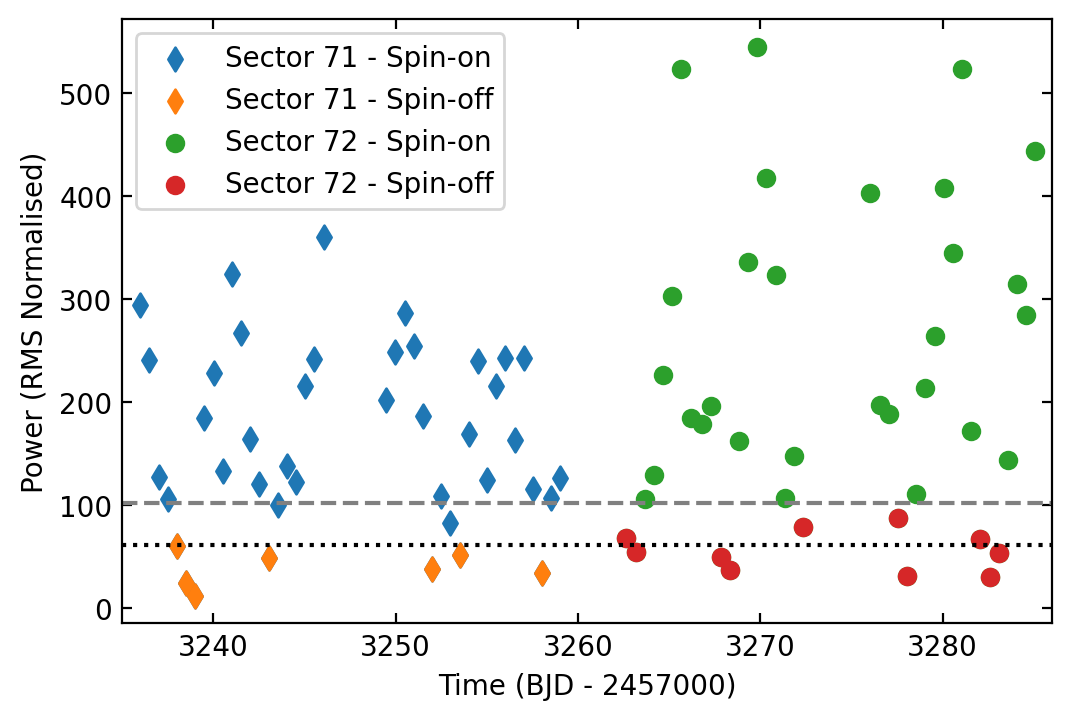}
    \caption{Time dependent evolution of power at $P_{\rm spin}$ for DW Cnc \tess\ data of sectors 71 and 72. The dotted line indicates the threshold for sector 71 at which the spin power is considered not significant. Similarly, the dashed line indicates a significance threshold for sector 72.}
    \label{fig:PSDLC}
\end{figure}
backs
To quantify the "on" and "off" states, we measure the power at $P_{\rm spin}$ for each of the 0.5-day segment for sectors 71 and 72 from Figure \ref{fig:dynPSD}. The resulting time series of the $P_{\rm spin}$ variability is shown in Figure \ref{fig:PSDLC}. From Figure \ref{fig:PSDLC} the Spin-off state is hence defined as when the $P_{\rm spin}$ power is below twice the average power at frequencies below $P_{\rm spin}$. These thresholds are indicated in Figure \ref{fig:PSDLC} by a dotted line for sector 71 and a dashed line for sector 72.

\begin{figure}
	\includegraphics[width=1\columnwidth]{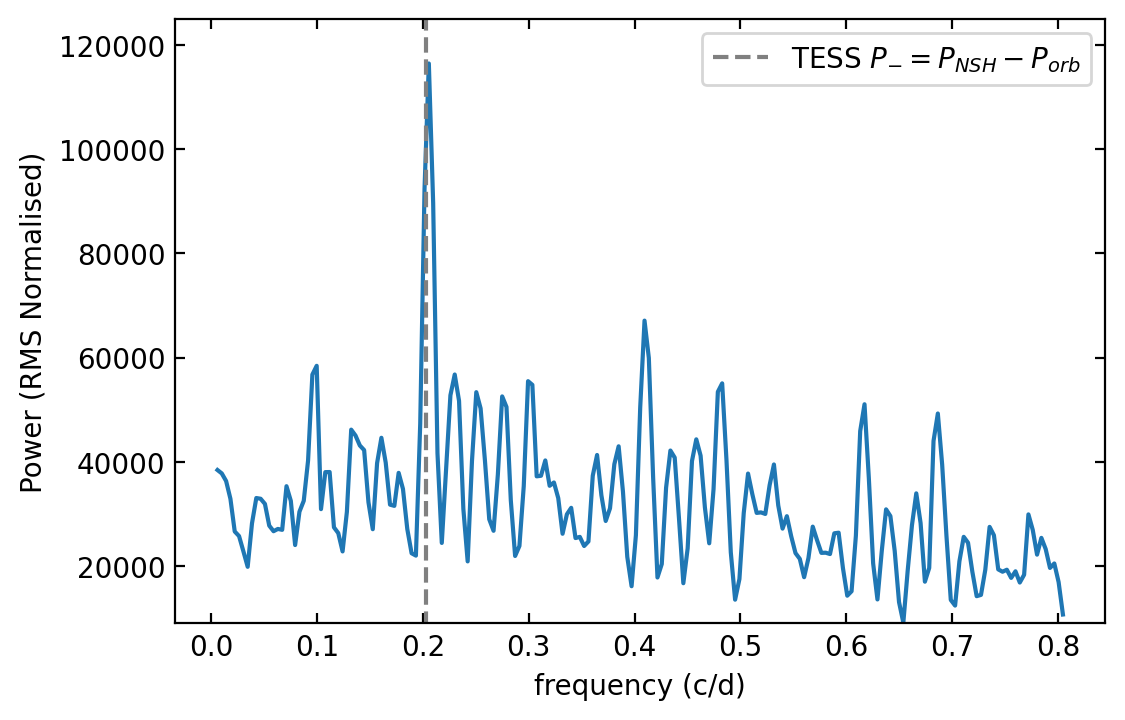}
    \caption{Lomb-Scargle periodogram of the time dependent variations of power at $P_{\rm spin}$ as shown in Figure \ref{fig:PSDLC}. The dashed line corresponds to the frequency inferred for the precession fundamental precession frequency associated with the negative superhump reported in Table \ref{tab:PSD}.}
    \label{fig:PSDLCPSD}
\end{figure}

Figure \ref{fig:PSDLCPSD} shows a Lomb-Scargle periodogram of the time series, to search for periodic modulation of the spin power (the data shown in Figure \ref{fig:PSDLC}). The PSD shows shows a single peak at $\sim$ 0.2 c d$^{-1}$, which is consistent with the fundamental precession frequency associated with the negative superhump observed in \tess\ data $1 / P_{\rm -} = 1 / P_{\rm nSH} - 1 / P_{\rm orb} = 0.2028(9)$c d$^{-1}$. The error on the signal is determined using the same method as described in Section \ref{ss:tess} for the signals in \tess\ data.


\begin{figure*}
	\includegraphics[width=1.0\textwidth]{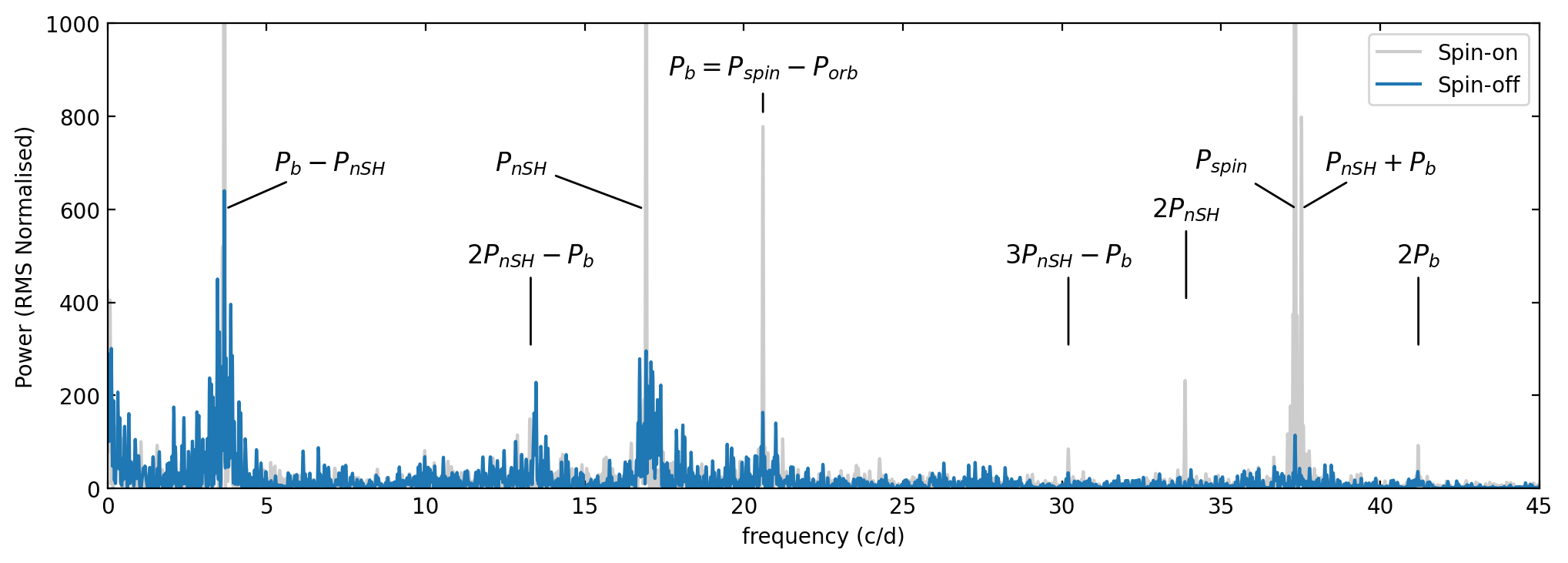}
    \caption{Lomb-Scargle power spectra of the separate ``on'' and ``off'' light curves from \tess\ sectors 71 and 72. The signals present in these sectors are marked as in Figure \ref{fig:PSD}.}
    \label{fig:on/offPSD}
\end{figure*}

\begin{figure}
	\includegraphics[width=1\columnwidth,trim={0 2cm 0 0}]{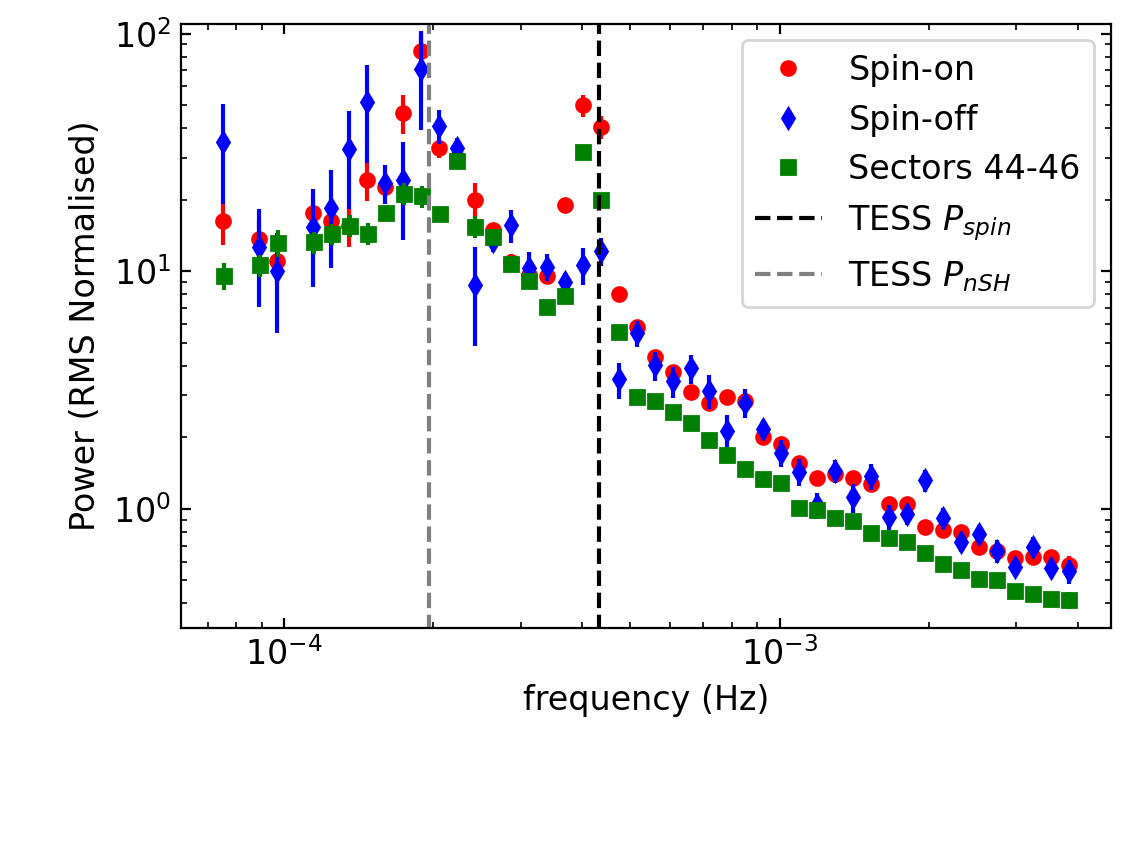}
    \caption{Time averaged power spectrum of \tess\ data. The squares correspond to Sectors 44 to 46, whilst sectors 71 and 72 are divided into the ``on'' (circles) and ``off'' (diamonds) spin state based on Figure \ref{fig:PSDLC}. The positions of $P_{\rm spin}$ and ${\rm nSH}$ from Table \ref{tab:PSD} are indicated by the dashed lines.}
    \label{fig:TPS}
\end{figure}

We also separated the light curves of sectors 71 and 72 into the "on" and "off" states based on the thresholds in Figure \ref{fig:PSDLC}, producing distinct light curves with and without the spin variability. The resulting Lomb-Scargle of these light curves are shown in Figure \ref{fig:on/offPSD}. We see only negligible power in the "off" state and a significant decrease in power amplitude at other frequencies present in the overall Sector 71 and 72 (see Figure \ref{fig:PSD} top panel). 

Furthermore, a time-averaged power spectrum (TPS) is computed for several sections of \tess\ data and shown in Figure \ref{fig:TPS}. The TPS is constructed by separating the light curve into 0.5-day segments. A segment of 0.5 days is chosen to be consistent with the segment length in Figure \ref{fig:dynPSD}. A Lomb-Scargle of each of the segments is then computed, before the individual PSDs are averaged together and binned onto a coarser frequency bin ($N_{\rm bins}=50$).

\begin{table}
    \centering
    \caption{\opticam\ observations corresponding to Table \ref{tab:opticam_log}. $f_{\rm low}$ represents the lowest possible frequency probed by the observation, defined as $1/effective~exposure$ from Table \ref{tab:opticam_log}. The following columns indicate if the spin period $P_{\rm spin}$ and the negative superhump $P_{\rm nSH}$ were detected in the given observation.}
    \label{tab:opticam_res}
    \begin{tabular}{lccr}
    \hline
    Date (UT) & $f_{\rm low}$ (c d$^{-1}$) & $P_{\rm spin}$ & $P_{\rm nSH}$ \\
    \hline
     2022-12-10  & 10.81 & \checkmark &  x \\
     2022-12-15  & 8.45 & x &  x \\
     2023-02-05  & 5.24 & \checkmark &  x \\
     2023-02-06  & 5.22 & \checkmark &  x \\
     2023-02-10  & 4.30 & \checkmark &  x \\
     2023-02-14  & 3.23 & x &  \checkmark \\
     2024-01-10  & 2.99 & \checkmark &  \checkmark \\
     2024-01-12  & 4.00 & \checkmark &  \checkmark  \\
     2024-01-14  & 3.17 & \checkmark &  \checkmark  \\
     \hline
    \end{tabular}

\end{table}

\begin{figure*}
	\includegraphics[width=1.0\textwidth]{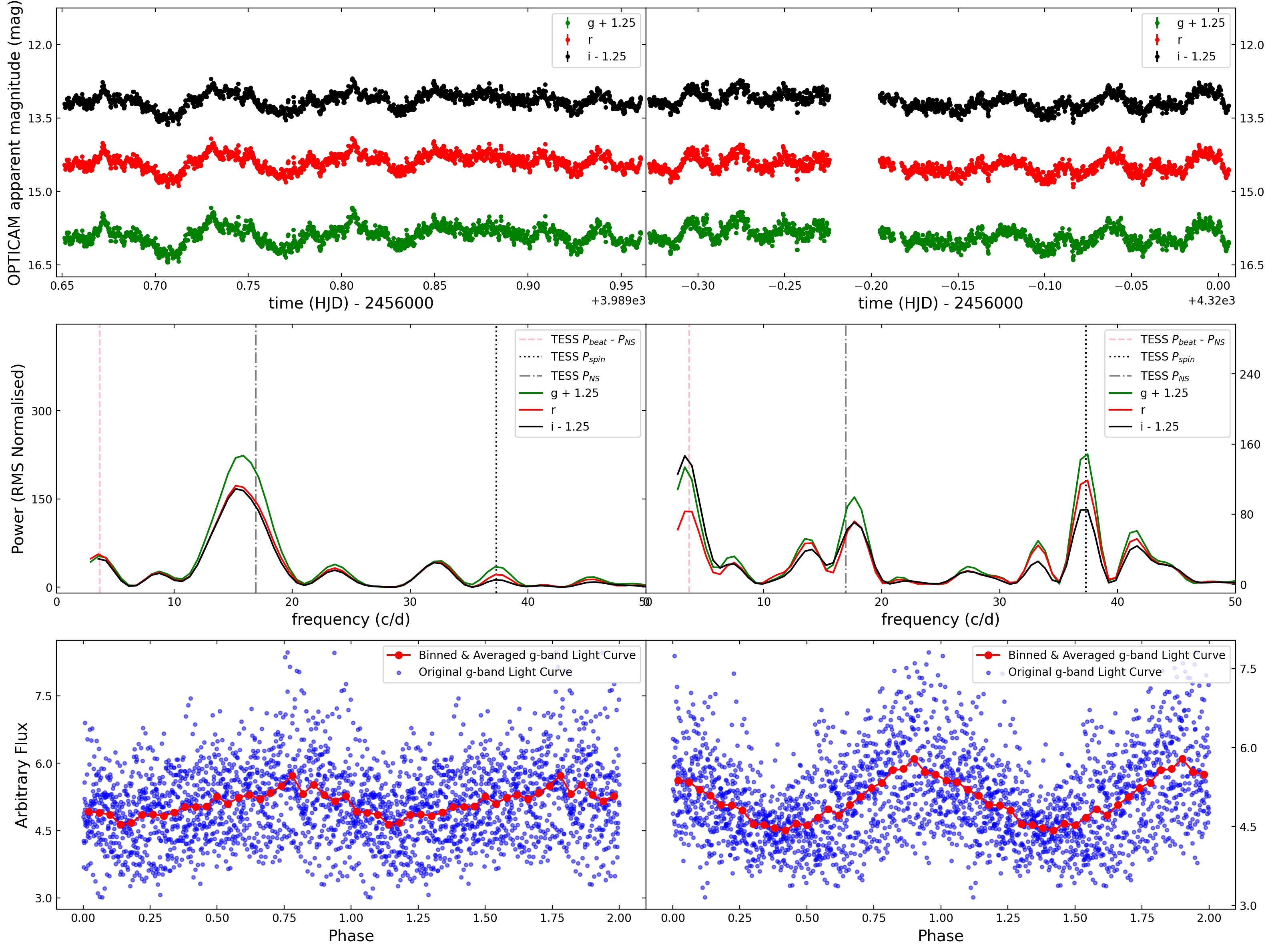}
    \caption{Examples of \opticam\ data from 2023-02-14 (left panels) and 2024-01-10 (right panels). The \textit{Top}: The top panels show a g, r and i band light curves of DW Cnc on given nights with a specified offset. \textit{Middle}: The middle panel shows a PSD corresponding to the light curves from the top panel with selected periods from \tess\ ($P_{\rm spin}$, $P_{\rm nSH}$ and $P_{\rm b} - P_{\rm nSH}$). \textit{Bottom}: The bottom panel shows a phase-folded $g$-band light curve, folded on the spin period $P_{\rm spin}$ from \tess\ data as noted in Table \ref{tab:PSD}.}
    \label{fig:opticamdata}
\end{figure*}

\begin{figure}
	\includegraphics[width=1\columnwidth,trim={0 4cm 0 0}]{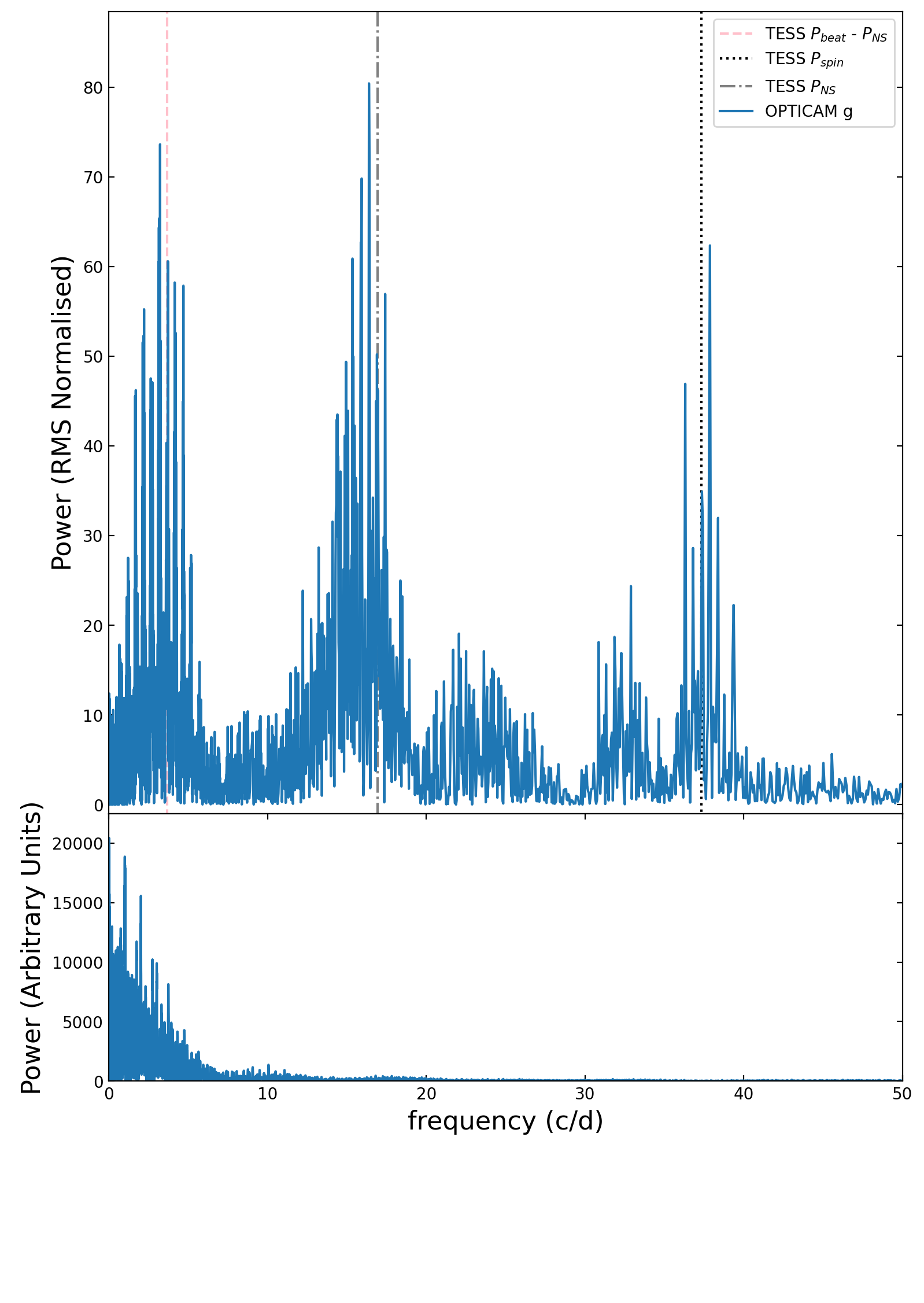}
    \caption{\textit{Top:} PSD of the combined \opticam\ $g$-band data from Table \ref{tab:opticam_log}. The vertical lines indicate the spin $P_{\rm spin}$, negative superhump $P_{\rm nSH}$ and the beat between the spin-orbit beat and the negative superhump $P_{\rm b} - P_{\rm nSH}$. \textit{Bottom:} PSD representing the window function used.}
    \label{fig:opticamdata_all}
\end{figure}

To estimate further constraints on the spin variability seen in \tess\ data of DW Cnc we introduce the \opticam\ data as detailed in Section \ref{ss:opticam}. The 3-band optical photometry was taken on multiple nights at different cadence and lengths (see Table \ref{tab:opticam_log} for further details). An example of the multi-band light curves is shown in the top panels of Figure \ref{fig:opticamdata}. In Figure \ref{fig:opticamdata} we choose to show examples of DW Cnc light curves from Table \ref{tab:opticam_log}, with the results from all the data being detailed in Table \ref{tab:opticam_res}.

Figure \ref{fig:opticamdata_all} shows the combined PSD of the \opticam\ data from Table \ref{tab:opticam_log}. The PSD is obtained as described in detail in Section \ref{ss:tess}. The signals in \tess\ data are marked by vertical lines, confirming the presence of spin of DW Cnc, the negative superhump and the beat between the spin-orbit beat and the negative superhump. To verify if all of these signals are always present in the \opticam\ data, the Lomb-Scargle is computed for all \opticam\ data and bands, with examples shown in the middle panels of Figure \ref{fig:opticamdata}. Given the shorter duration of the individual \opticam\ observations and the limiting lowest frequency, we use the frequencies in Table \ref{tab:PSD} as guidelines to mark the likely presence of signals in the \opticam\ data. The presence of the signals detected in Figure \ref{fig:opticamdata_all} in separate \opticam\ observations is marked in Table \ref{tab:opticam_res}. We note that the spin signal wasn't detected during several observations, similar to the behaviour seen in \tess\ sectors 71 and 72.

The 2 columns in Figure \ref{fig:opticamdata} represent examples of when the spin signal is not present (left) and where is it detected (right). There is also a clear power excess seen at the frequencies corresponding to the negative superhump frequency from 14$^{\rm{th}}$ of February 2023 (almost a year before sector 71) until the last measurement on the 14$^{\rm{th}}$ of January 2024 ($\sim$ month after the end of sector 72). To demonstrate the switching of spin between the ``on'' and ``off'' state is present in the \opticam\ data as well, phase-folded light curves of $g$-band \opticam{} data are included in the bottom panels in Figure \ref{fig:opticamdata}. There, the \opticam{} data is folded on the \tess\ spin as shown in Table \ref{tab:PSD}.

\subsection{Bursts}
\label{ss:bursts}

The long-term \asassn\ and \atlas\ light curve of DW Cnc covers its variability from 16$^{\rm{th}}$ of February 2012. There are only a couple of instances in this time when the luminosity of DW Cnc rises above its average levels, showing distinct bursts akin to those reported in \citet{Crawford2008JAVSO..36...60C}. At $\sim$ 2456600 BJD there are 2 data points at $\sim$ 3 $\times$ 10$^{32}$ erg~s$^{-1}$ in \asassn\ $V$-band. However, in both cases, these are single-point measurements and considering the uncertainty of the measurements the validity of these points as an outburst may be less reliable.

The data show two low-amplitude outbursts that we will not consider further here. These appear immediately before the low state between $\sim$ 2458000 BJD and 2459000 BJD \citep{Covington2022ApJ...928..164C} (shaded region in Figure \ref{fig:LC}), at 2458164 BJD reaching peak luminosity at $\sim$ 8.8 $\times$ 10$^{32}$ erg~s$^{-1}$ lasting for about a day in \asassn\,. However, \atlas\ observed DW Cnc in quiescence in the middle of the \asassn\ bursts. Therefore, the outburst is either not real or a sequence of multiple $\sim$0.5 day bursts. In any case, this is not considered here due to the low significance. A similar amplitude burst is seen in \asassn\ at $\sim$  2460287 BJD with peak luminosity of $\sim$ 1.3 $\times$ 10$^{33}$ erg~s$^{-1}$ and also about a day long duration.

However, a truly analogous burst to those reported in \citet{Crawford2008JAVSO..36...60C} is seen in \asassn\ at HJD 2460450. The peak luminosity of this burst is recorded at 6.6 $\times$ 10$^{33}$ erg~s$^{-1}$, corresponding to a brightening of 4 magnitudes from its quiescent levels. By combining the \asassn\ and \atlas\ light curves it is possible to obtain an upper limit on the duration of the burst;  Due to the \asassn\ light curve sampling, the upper limit on the duration of the burst would be $\sim$ 6 days, when a measurement at quiescent level confirms the end of the burst after a $\sim$ 4-day data gap. However, considering the \atlas\ $c$-band measurements as seen in the right inset plot in Figure \ref{fig:LC}, the upper limit on the duration is $\sim$3 days. The burst has been completely missed by \atlas{} and only serendipitously picked up by \asassn{}. This raises an important question, how many similar shorter-length bursts have been missed in DW Cnc since the start of monitoring and in other systems as well? Therefore, with the rise of higher-cadence surveys such as \blackgem{} and the Rubin Observatory \citep{LSST2019ApJ...873..111I}, the frequency of detecting the short-duration bursts in AWDs should allow us to understand them better and constrain their recurrence with greater precision. 

The total energy released during the burst is determined by integrating the luminosity under the light curve after subtracting the baseline luminosity of the light curve. The baseline is determined from a fit to the mean of the pre-burst long-term light curve about $\sim$ 100 days before the burst. This reveals an energy release of $>6\times\,10^{38}\,\mathrm{erg}$ when using a linear interpolation between the first quiescent point before and after the burst from the combined \asassn\ and \atlas\ light curve. A lower limit here is used as no bolometric correction is used and energy that may have been released in other parts of the spectrum has not been considered. A more conservative lower limit, only integrating the energy of the points above the quiescence level, would produce an energy of $>9\times\,10^{37}\,\mathrm{erg}$.

\section{Discussion}
\label{s:disc}

In Section \ref{ss:micronova} we discuss the interpretation of the bursts in DW Cnc as a micronova, as well as the related implications with this interpretation. In Section \ref{ss:signals} we discuss the interpretation of signals detected in \tess\ data of DW Cnc with the physical and geometrical consequences for the system. Section \ref{ss:pole_flipping} details the interpretation of the spin variability seen in DW Cnc.

\subsection{Burst Nature}
\label{ss:micronova}

There are several possible explanations for bursts observed in magnetic AWDs, such as dwarf nova outbursts, micronovae, and magnetically gated bursts. A diagnostic diagram that attempts to distinguish among these possibilities has been introduced by \citet{Ilkiewicz2024ApJ...962L..34I}, where the burst integrated energy, peak luminosity, recurrence time and burst duration appear to separate different burst types. Adopting a peak luminosity for the burst observed in DW Cnc of $\approx$6.6 $\times$ 10$^{33}$ erg~s$^{-1}$, an integrated energy of $>6\times10^{38}$ erg, and a bust duration of $<$ 3 days, it appears that the burst in DW Cnc most closely matches the properties of ASASSN$-$19bh \citep{Scaringi2022} and PBC J0801.2$-$4625 \citep{Irving2024MNRAS.530.3974I} which have both been interpreted as a particularly energetic micronovae events. 

As discussed in \citet{Scaringi2022a} micronovae may occur if mass accretion occurs onto magnetically confined polar regions of AWDs, where accreted mass is magnetically confined and accumulates until thermonuclear ignition conditions are reached, triggering a runaway micronova explosion. Under these assumptions the recurrence time of micronova events on any one system can be estimated from the mass accretion rate $\dot{M}_{acc}$ together with the total mass required to thermonuclear ignite the magnetically confined accretion column. 

Adopting a mass to energy conversion rate of hydrogen burning during the CNO cycle of $\approx$10$^{16}$ erg g$^{-1}$ \citep{Bode2008,Jose2020}, the observed $>6 \times 10^{38}$ erg burst in DW Cnc would convert to $M_{\rm col} >3 \times 10^{-11} \rm M_{\rm \odot}$. Assuming this amount of mass has been accumulated onto a magnetically confined polar region since the last burst an estimate of the mass accretion rate onto DW Cnc can be obtained from $\dot{M}_{\rm acc} = {M_{\rm col} / t_{\rm rec}}$. The $M_{\rm col}$ represents the mass that is ejected from the accretion column during the burst and $t_{\rm rec}$ the recurrence time of the bursts. The \asassn\ light curve in Figure \ref{fig:LC} reveals a recorded (small amplitude) burst $\approx 6$ years earlier, which we assume to be a partially observed micronovae event to obtain a mass accretion rate onto DW Cnc of $5\times 10^{-12} \rm M_{\rm \odot}$yr$^{-1}$.

This estimate can be compared to the mass transfer rate of DW Cnc assuming the optical and X-ray luminosities are a reliable tracer for accretion luminosity through
\begin{equation}
    L_{\rm acc} = \frac{G M_{\rm WD} \dot{M}}{2 R_{\rm in}} \approx L_{\rm opt} + L_{\rm X-rays}
\end{equation}
\noindent where $M_{\rm WD}$ is the WD mass, assumed to be $\sim$ 0.8 $M_{\rm \odot}$, $\dot{M}$ is the mass transfer rate, $G$ the gravitational constant and $R_{\rm in}$ the inner disc radius. We set the inner disk radius to the WD radius, $R_{\rm in} = R_{\rm WD} = 0.01 R_{\rm \odot}$ as any X-ray emission from accretion curtains should also be considered. Adopting the optical luminosity $L_{\rm opt} = 9.8 \times 10^{31}$ erg~s$^{-1}$ from the \asassn\ light curve (excluding bursts), together with the X-ray luminosity of $L_{\rm X} = 8.7 \times 10^{31}$ erg~s$^{-1}$ in the \xmm\ 0.3 $-$ 10 keV range as inferred by \citet{Nucita2019MNRAS.484.3119N}, yields a mass transfer rate $\dot{M} \approx 3.9 \times 10^{-11}$ $M_{\rm \odot}$yr$^{-1}$, consistent with other IPs at similar orbital periods \citep{Knigge2011,Duffy2022MNRAS.510.1002D}.

It is important to note that while the the optical and X-ray luminosities provide an estimate of the secular disk mass \textit{transfer} rate, the burst energetics provide an estimate of mass \textit{accretion} rate onto the WD magnetically confined polar regions. Although these can be the same in some cases, they do not strictly have to be. As also discussed in \cite{Scaringi2022a} the mass transfer and mass accretion rates would not necessarily be the same since some of the material may experience either lateral spreading at the base of the magnetically confined polar region or be accreted outside of the magnetically confined region (or both). In any case we emphasise that the interpretation of \cite{Scaringi2022a} and related methodology applied here to DW Cnc requires further testing through observational evidence and theoretical modelling.

\subsection{\tess\ Signals}
\label{ss:signals}

The Lomb-Scargle of the \tess\ light curves as presented in Figure \ref{fig:PSD} shows many periodic signals reported in Table \ref{tab:PSD}. Only 3 signals are always present in the \tess\ data. One of them, at 37.32166(7) c d$^{-1}$ is consistent with the spin period of the systems previously reported \citep{Rodriguez-Gil2004MNRAS.349..367R}. This confirms what was found in \citet{Covington2022ApJ...928..164C}, i.e. that the spin signal has recovered after its disappearance in the low state \citep{Montero2020MNRAS.494.4110S}. Another period that is always present in the \tess\ light curve is the 20.5962(2) c d$^{-1}$ signal. This signal has been observed before and after the low state \citep{Montero2020MNRAS.494.4110S,Covington2022ApJ...928..164C} and is associated with the beat between the spin and the orbital period of the system. The orbital period is not recovered due to the likely low inclination of the system. The last signal present during all \tess\ sectors is the first harmonic of the beat, at 41.20 c d$^{-1}$. 

Sectors 71 and 72 see the appearance of new signals. One of them is the 16.9276(2) c d$^{-1}$ signal. This is very close to the spectroscopic period of the system at 16.72441(6) c d$^{-1}$ \citep{Montero2020MNRAS.494.4110S}. Given the close frequencies of the signals, we propose that 16.9276(2) c d$^{-1}$ is the negative superhump of the system associated with the retrograde precession of a tilted accretion disc. This is similar to another low inclination system MV Lyr, which also displays negative superhump \citep{Bruch2023MNRAS.519..352B}. No precession frequency has however been recovered. TV Col, another intermediate polar showing micronova bursts, has also shown negative superhump \citep{Scaringi2022}; and even showed a positive superhump in the pre-burst stage. This peculiar behaviour suggests that DW Cnc may also show similar behaviour before undergoing a burst and thorough monitoring is necessary to confirm this for future bursts. 

The other signals represent beats with the harmonics of the superhump and the spin-orbit beat, as described in Table \ref{tab:PSD}.

\subsection{Spin variability as evidence of pole flipping?}
\label{ss:pole_flipping}

As described in Section \ref{ss:spin} and shown in Figure \ref{fig:dynPSD} the spin period is showing semi-periodic changes between "on" and "off" states in \tess\ sectors 71 \& 72, with evidence that the behaviour started as early as February 2023 from the \opticam\ observations. The change between states occurs on 0.2028(9) c d$^{-1}$ timescale, consistent with the fundamental superorbital period expected at $\frac{1}{P_{\rm -}} = \frac{1}{P_{\rm nSH}} - \frac{1}{P_{\rm orb}} = 0.2028$ c d$^{-1}$ $\sim$ 4.9 days, as seen in Figure \ref{fig:dynPSD} and \ref{fig:PSDLCPSD}. Unfortunately, this period is not detected in the \tess\ data but can be recovered from the variability of power at $P_{\rm spin}$. 
One possible explanation for this behaviour is a modulation of accretion, where accretion ceases during the "off" states. This would resemble the low state previously observed in DW Cnc \citep{Montero2020MNRAS.494.4110S} but occurring on much shorter timescales. However, in contrast to \citet{Montero2020MNRAS.494.4110S}, no significant drop in flux is observed, suggesting that accretion persists even during the "off" states. This raises the possibility that the modulation is related to accretion geometry rather than an outright cessation of accretion.

The hypothesis that pole flipping occurs in DW Cnc must be treated with caution. In asynchronous polars (APs), pole flipping is driven by accretion from an azimuthally asymmetric stream directly interacting with the white dwarf's magnetosphere \citep{Ferrario1999MNRAS.309..517F}. This scenario is known to occur in APs, where the central white dwarf accretes via magnetically confined stream, but the orbit and spin period are slightly out of sync. In these systems, particularly BY Cam and CD Ind \citep{Littlefield2019ApJ...881..141L}, where both accretion poles are visible, the switch from one to another can be detected by a phase shift of the spin during the beat cycle. In APs, pole-switching is accompanied by a phase shift of $180^{\circ}$ in the spin cycle during half the spin-orbit beat cycle.  This is observed in FO Aqr \citep{Norton1992MNRAS.254..705N} as well as CD Ind \citep{Littlefield2019ApJ...881..141L} and BY Cam \citep{Mason2022ApJ...938..142M}. Such behaviour is unlikely in DW Cnc, where the presence of a negative superhump implies accretion via a tilted, precessing accretion disc. Unlike APs, where pole flipping occurs at the beat frequency between the spin and orbital periods, any pole modulation in DW Cnc would likely occur at the disc's precession period. This distinction underscores that pole flipping in DW Cnc, if present, would be fundamentally different from that in APs.

The observed spin variability may instead reflect changes in the accretion flow alignment with the white dwarf's magnetic poles, modulated by the disc's tilt and precession. Figure \ref{fig:2dprofiles} shows a similar phase shift of $\sim180^{\circ}$ in the beat during the negative superhump cycle and in the spin phase during the precession cycle, suggesting that the tilt and precession of the disc influence the accretion flow. Furthermore, before the discontinuous jump of spin phase at $\sim$ 0.3 and $\sim$ 1.6 precession phase there is a continuous shift in the spin over $\sim$ 0.5 cycle of the precession. This modulation could lead to periodic shifts in the dominant accreting pole. Unlike classical pole flipping, which depends on direct interaction with a stream, this process would be driven by the precession of the disc itself, which modifies the effective area of interaction between the disc and the magnetic field. Furthermore, the process needs not to be discrete, but rather a continuous change of accretion power fraction between the magnetic poles, as indicated by the right panel in Figure \ref{fig:2dprofiles}. 

The non-detection of a photometric modulation at the orbital period suggests the inclination is fairly low.  If that is the case, the "off" state may correspond to either accretion onto only one magnetic pole permanently obscured by the WD, or a gradual change of power accreted onto either pole. Such a gradual change may be supported by Figure \ref{fig:2dprofiles} and in the remnants of sinusoidal variations in phase folded light curves of the spin in the "off" state in \tess\ data (Figure \ref{fig:dynPSD}) and \opticam\ data (Figure \ref{fig:opticamdata}). These phase-folded light curves on the \tess\ spin period show a single peaked shape consistent with majority accretion onto a single magnetic pole. Furthermore, there is no expectation of the geometry of the accretion disc changing, as seen in Figure \ref{fig:TPS}. This figure shows, that the broad-band shape of the PSD in DW Cnc is the same during "on" and "off" states as well as prior to the appearance of the negative superhump. Considering the relatively red bandpass of \tess\ data, this is not surprising as the bandpass would be only probing the outer edges of the accretion disc. \citet{Nucita2019MNRAS.484.3119N} observed DW Cnc with \xmm\ in 2012 and found that the spin signal is present, and shows no energy dependence, further suggesting that only one pole is visible. 



Using \opticam\ data, we can further constrain that the negative superhump was present in the system almost a year prior to \tess\ sector 71. With the lack of spin in the \opticam\ data from 14$^{\rm{th}}$ of February 2023, during which the negative superhump is detected for the first time, it may be possible to assume that the spin variability and negative superhump are coupled. However, the only other instance in which the spin is not detected (15$^{\rm{th}}$ of December 2022) is before the appearance of the negative superhump, this is most likely due to the lower data quality and substantial gaps in the dataset. Furthermore, the \opticam\ data suggests a continuation of this behaviour after the end of Sector 72 for at least another month, up to 4 months before the 4 magnitude burst.

A hypothetical scenario may then suggest that once a sufficient amount of material has been accreted on one pole a micronova burst can be triggered. Assuming that similar behaviour occurs before a burst, the recurrence time during which material actively accretes onto one magnetic pole to trigger a micronova is at least half of the time between the observed bursts. However, without at least a very precise recurrence timescale of the bursts, this cannot be verified. Similar IPs, such as EX Hya could therefore eventually exhibit such behaviour. With similar amplitude bursts reported \citep{Bond1987IBVS.3037....1B,Hellier1989MNRAS.238.1107H,Hellier2000MNRAS.313..703H}, EX Hya is a prime candidate for monitoring to find another system with variable spin and test the relation of its variability as pre-cursor of micronova bursts.

\section{Conclusions}
\label{s:conslusion}

DW Cnc is a well-studied intermediate polar which has been known to show bursts of unexplained nature in the past. Here we present new \tess\ and ground-based \asassn{} \atlas{} and \opticam{} data of DW Cnc. We report a new 4-magnitude burst in the \asassn{} light curve, which we interpret as a micronova burst. We present the similarities of DW Cnc behaviour with other magnetic systems showing micronovae as well as categorising it using the diagnostic diagrams from \citet{Ilkiewicz2024ApJ...962L..34I}. We also report the appearance of a negative superhump in 2 of the 5 \tess\ sectors which we retrospectively confirm has appeared at least a year earlier in the \opticam\ data. Along with the negative superhump we report a variability in the spin amplitude on the precession period associated with the negative superhump. We do not observe any changes in the geometry of the accretion disc or flux levels during the times at which the spin amplitude is negligible. However, we observe a decrease in the amplitude of all the other signals present in the PSD. We associate this behaviour to several possible scenarios, where in the "off" state the accretion mostly occurs onto the accretion pole obscured by the WD. An observed $\sim180^{\circ}$ shift in the spin phase over the precession cycle of the disc as well as $\sim180^{\circ}$ shift in the beat phase over the negative superhump cycle further suggests that the accreting poles are switching and that this phenomena is linked to the tilt of the accretion disc. We are not able to distinguish this unequivocally, but we propose that the geometry could be verified with fast time-resolved spectroscopy to separate "on" and "off" states. 

\section*{Acknowledgements}

This paper includes data collected with the TESS mission, obtained from the MAST data archive at the Space Telescope Science Institute (STScI). Funding for the TESS mission is provided by the NASA Explorer Program. STScI is operated by the Association of Universities for Research in Astronomy, Inc., under NASA contract NAS 5–26555. MV acknowledges the support of the Science and Technology Facilities Council (STFC) studentship ST/W507428/1. SS is supported by STFC grants ST/T000244/1 and ST/X001075/1. Z.A.I acknowledges support from STFC grant ST/X508767/1. This work has made use of data from the Asteroid Terrestrial-impact Last Alert System (ATLAS) project. The Asteroid Terrestrial-impact Last Alert System (ATLAS) project is primarily funded to search for near-earth asteroids through NASA grants NN12AR55G, 80NSSC18K0284, and 80NSSC18K1575; byproducts of the NEO search include images and catalogues from the survey area. This paper is based upon observations carried out at the Observatorio Astron\'omico Nacional on the Sierra San Pedro M\'artir (OAN-SPM), Baja California, M\'exico, which is operated by the Universidad Nacional Aut\'onoma de M\'exico (UNAM). JE would like to acknowledge the PAPIIT funding project IN-113723. This work was partially funded by Kepler/K2 grant J1944/80NSSC19K0112 and HST GO-15889, and STFC grants ST/T000198/1 and ST/S006109/1.  AC acknowledges support from the Newton International Fellowship grants NF/170803 and AL/221034. The ATLAS science products have been made possible through the contributions of the University of Hawaii Institute for Astronomy, the Queen’s University Belfast, the Space Telescope Science Institute, the South African Astronomical Observatory, and The Millennium Institute of Astrophysics (MAS), Chile. P.J.G. is partly supported by the National Research Foundation of South Africa, through SARChI grant 111692. DdM acknowledges support from INAF Programma di Ricerca Fondamentale. DAHB acknowledges support from the National Research Foundation. We thank Scott Hagen for the helpful discussions on accretion geometry.

\section*{Data Availability}

The \tess\ data used in the analysis of this work is available on the MAST webpage \url{https://mast.stsci.edu/portal/Mashup/Clients/Mast/Portal.html}. The \asassn\ data \citep{Shappee2014,Kochanek_2017} used for the calibration of \tess\ data is available on the \asassn\ webpage \url{https://asas-sn.osu.edu/}.



\bibliographystyle{mnras}
\bibliography{refs} 





\bsp	
\label{lastpage}
\end{document}